\documentclass[aps,twocolumn,groupedaddress,showpacs]{revtex4}
\usepackage{epsf,latexsym}
\usepackage{times}

\newcommand{\bra}[1]{\langle #1 |}
\newcommand{\ket}[1]{| #1 \rangle}

\newcommand\R{{\mathrm {I\!R}}}

\newcommand{\be}{\begin{equation}}
\newcommand{\ee}{\end{equation}}
\newcommand{\bae}{\begin{eqnarray}}
\newcommand{\eae}{\end{eqnarray}}

\def\CC{{\rm\kern.24em \vrule width.04em height1.46ex depth-.07ex
    \kern-.30em C}}
\def\P{{\rm I\kern-.25em P}}

\def\bbbc{{\mathchoice {\setbox0=\hbox{$\displaystyle\rm C$}\hbox{\hbox
to0pt{\kern0.4\wd0\vrule height0.9\ht0\hss}\box0}}
{\setbox0=\hbox{$\textstyle\rm C$}\hbox{\hbox
to0pt{\kern0.4\wd0\vrule height0.9\ht0\hss}\box0}}
{\setbox0=\hbox{$\scriptstyle\rm C$}\hbox{\hbox
to0pt{\kern0.4\wd0\vrule height0.9\ht0\hss}\box0}}
{\setbox0=\hbox{$\scriptscriptstyle\rm C$}\hbox{\hbox
to0pt{\kern0.4\wd0\vrule height0.9\ht0\hss}\box0}}}}

\def\bbbz{{\mathchoice {\hbox{$\sf\textstyle Z\kern-0.4em Z$}}
{\hbox{$\sf\textstyle Z\kern-0.4em Z$}}
{\hbox{$\sf\scriptstyle Z\kern-0.3em Z$}}
{\hbox{$\sf\scriptscriptstyle Z\kern-0.2em Z$}}}}

\begin{document}
\title{Universal quantum control in irreducible state-space sectors:
\\application to bosonic and spin-boson systems}
\author{Paolo Giorda$^{1,3}$  Paolo Zanardi$^{1,2,3}$  and Seth Lloyd$^2$
}
\affiliation{$^1$ Institute for Scientific Interchange (ISI), Villa Gualino, Viale 
Settimio Severo 65, I-10133 Torino, Italy}

\affiliation{$^2$ Department of Mechanical Engineering,
Massachusetts Institute of Technology, Cambridge Massachusetts 02139
}
\affiliation{$^3$  Istituto Nazionale per la Fisica della Materia (INFM), UdR Torino-
Politecnico, 10129 Torino, Italy}

\begin{abstract}
We  analyze the dynamical-algebraic approach to universal quantum control  introduced in P. Zanardi, S. Lloyd, quant-ph/0305013.
The quantum state-space $\cal H$ encoding information decomposes into irreducible sectors and subsystems associated to the group 
of available evolutions. If this group coincides with the unitary part of the group-algebra $\CC{\cal K}$ of some  group 
$\cal K$ then universal control is achievable over the
${\cal K}$-irreducible components of $\cal H$. This general strategy is applied to different kind of bosonic systems.
We first consider massive bosons in a double-well and show how to achieve universal control over all  finite-dimensional
 Fock sectors. We then discuss a multi-mode massless  case
giving the conditions for generating the whole infinite-dimensional multi-mode Heisenberg-Weyl enveloping-algebra.
Finally we show how to use an auxiliary bosonic mode coupled to finite-dimensional systems to generate 
high-order non-linearities needed for universal control.
\end{abstract}

\pacs{ }
\maketitle
\section{introduction}
Over the last few years  quantum information science taught us how to take advantage
of information encoded  in the states of a quantum system \cite{QC}.
A fundamental requirement is then the ability to process such information 
in the most general fashion by resorting to the physically available interactions.
When arbitray transformations (or approximations of) over the system state space $\cal H $ are dynamically realizable 
one says that {\em universality} is achieved.
To this end one typically introduces a set of elementary building blocks,
e.g., quantum logic gates, whose combinations generate the whole set of desired dynamical evolutions.
A very important example is provided by {\em quantum computation} \cite{QC}.
There the prototype information-encoding system is provided by a collection of two-level systems
i.e., {\em qubits}, such that each of those subsystems is completely controllable i.e.,
any $SU(2)$ rotation is realizable, and an entangling  two-qubit transformation is available
\cite{SETH},\cite{UG},\cite{mike}. 

Unfortunately there are several physically relevant situations e.g., spin-based quantum computing, 
in which  such a goal is extremely difficult to achieve practically. In those cases
the lack of easily available dynamical resources forces one to consider
more sophisticated ways of encoding and manipulating information. For example
it might be the case that independent single qubit-operations are very difficult to realize
in view of the small spatial separation between qubits. In this case
just global operations affecting the  quantum register as a whole  
are readily available;  these latter clearly do not  generate the
full set of transformation over $\cal H$. Another class of examples of this state
 of affairs is provided when  all naturally available interactions commute with some observable
e.g., total spin. It follows that the state-space splits in mutually  
inaccesible sectors corresponding to the different eigenvalues of such an observable.

This kind of problem led several authors to develop
the concept  of {\em encoded universality}  \cite{danFT,danNS,enc,Dan,lorenza}.
That was  mostly done  by explicitly working out a few important and non-trivial 
istances of a such a notion. Roughly speaking the idea is that even in presence of a limited set
of available control resources one can find out a in invariant  subspace $\cal C$ of $\cal H$
over which universality is achievable.  More recently a general  dynamical-algebraic
framework has been suggested to underlie all these former examples 
and to provide room for further generalizations  and applications  \cite{ZL-ctrl}.

The  key feature of  the techniques advocated in Ref.  \cite{ZL-ctrl}
is a massive use of tools borrowed from group representation theory \cite{corn}. 
Assuming that the set  evolutions one can enact forms a group i.e., the {\em control group}
${\cal U}_A,$ the state-space splits according the associated irreducible representations
(irreps). In such invariant subspaces ${\cal U}_A$  typically still does not
 allow for universal control, nevertheless that goal can be accomplished
when one has at disposal generic Hamiltonians belonging to the {\em group algebra} of ${\cal U}_A.$
It follows that if  one recognizes that the controllable interactions belong to the group-agebra
of some hidden non-abelian group $\cal K$ universal controllability is realizable over the whole family of subspaces 
and subsystems associated with the irreps of $\cal K$ \cite{ZL-ctrl}.
In this approach an important role is played by the use of {\em symmetries}
that allow one  to build dualities between different  kind of encodings and manipulation 
strategies. This crucial role is also at the basis of the deep conceptual connection
between universal controllability strategy  proposed in \cite{ZL-ctrl}
and the theory of quantum noise correction and avoidance \cite{EAC},\cite{KLV},\cite{stab}.

In this paper we shall further develop and apply  the general lines of Ref. \cite{ZL-ctrl}.
The relevant conceptual and mathematical framework will be illustrated in a fully detailed manner and
novel physical examples worked out. A major emphasis will be given to bosonic systems,
both massive ones (with associated finite dimensional irreducible sectors)  
and massless one (infinite dimensional Fock space).

The paper is organized as follows. 
In section II will be introduced the basic terminology
and quantum-control notions. In section III we will lay down
the general group-theoretic formalism at the basis of our 
approach. In section IV and V we will analyze the controllability
of a pair of  coupled massive bosonic modes in terms of the associated
$SU(2)$ structure. In section VI the massless bosonic case
with infinite-dimensional state space   will be considered.
In section. VII we will discuss an auxiliary bosonic mode technique for generating
highly-non linear interactions. Finally section VIII will contain
the conclusions.

\section{ Preliminaries}
The starting point is the set of physically available interactions that act over the quantum state-space 
${\cal H}$:
\be
{\cal I}_A:=\{ H(\lambda)\}_{\lambda\in {\cal M}}\subset \mbox{End}({\cal H})
\label{IA}
\ee
where $\cal M$ is the set of possible values of the control parameters $\lambda$.
The physically realizable control processes can be described by $\lambda (t)\in{\cal P}_A$, i.e. by
a set  of ${\cal M}$-valued functions (paths)  that represent the  evolutions of the
control parameters that  are experimentally achievable.
Thus the unitary quantum evolutions  one can enact are those that are governed by the time-dependent 
Hamiltonians $H(\lambda(t))$:
\be
U(\lambda)=T\exp ( -i\int_{\R}  H(\lambda(t)) dt ) \quad (\lambda\in {\cal P}_A).
\label{allowedU}
\ee
The {\em pair} $({\cal I}_A, {\cal P}_A)$ describes the physical resources available in the given 
experimental situation.

In the following we will assume that if $U$ is an allowed evolution, then $U^\dagger$ is allowed as well;
we also assume that the trivial evolution, i.e. $U=\openone,$ is an allowed one.
With this assumption the set  of unitary transformations that one can generate in the experimental situation
described by $({\cal I}_A, {\cal P}_A)$ has the structure of a {\em group} ${\cal U}_A$.

We now give a brief description of the properties of the group ${\cal U}_A$ that depend on the  paths 
${\cal P}_A$ that can be realized. We start by describing a well-know result of quantum control theroy.
Suppose that for any two arbitrary operators $A,B \ \in {\cal I}_A$ one is allowed to apply sequences of 
pulses $\exp(\pm iA\delta t), \exp(\pm iB\delta t)$, than the following relations hold
\be
e^{\pm iA\delta t}e^{\pm iB\delta t}= e^{i(\pm A \pm B) \delta t} + O(\delta t^2);
\label{trot1}
\ee
\be
e^{iA\delta t}e^{iB\delta t}e^{-iA\delta t}e^{-iB\delta t}= e^{i(AB-BA)\delta t^2} + O(\delta t^3).
\label{trot2}
\ee 
This means that if one can drive the control parameters along arbitrary
paths in $\cal M$, then one can generate, by iteratively applying the sequences (\ref{trot1}) and (\ref{trot2}),
all the possible linear combinations of Hamiltonians of the form $\pm iA, \pm i[A,B], \pm i [[A,B],C]$, ect.
, where $A,B,C,...\in {\cal I}_A$ \cite{SETH}. The result can be stated in terms of unitary operators as:
\begin{equation}
{\cal U}_A=e^{ {\cal L}_A },
\label{ctrl}
\end{equation}
where ${\cal L}_A$ is the set of Hamiltonian generated from the set  ${\cal I}_A$ by commutation.
${\cal L}_A$ it has the structure of Lie algebra.
On the contrary, if we suppose that there are restrictions on the experimentally realizable paths ${\cal P}_A $ 
then the allowed unitary transformations are such that
${\cal U}_A \subset e^{ {\cal L}_A }$.

From the point of view of the control theory the question that naturally arises is the following:
given the {\em pair} $({\cal I}_A, {\cal P}_A),$  what kind of control one has on the physical system $S$ under study?
One has {\em universal control} over $S$ when it is possible to enact all the unitary operators ${\cal U}({\cal H})$
where ${\cal H}$ is the quantum state-space of $S$. 
Now, depending on the {\em pair} $({\cal I}_A, {\cal P}_A)$ there can be different situations:
\begin{itemize}
\item{}
if ${\cal U}_A={\cal U}({\cal H})$ universal control over ${\cal H}$ is achieved;
\item{}
if  ${\cal U}_A$ is {\em dense} in  ${\cal U}({\cal H})$  then universal control over  ${\cal H}$ is still achieved
since it is possible to simulate any unitary evolution with arbitrary accuracy by means of the available  resources 
\item{}
if ${\cal U}_Ai$ is not dense  in ${\cal U}({\cal H}) $, then  it is not possible to have universal control over  ${\cal H}$; in this case one can
indtroduce the notion of {\em  encoded universality}.
\end{itemize}
When the {\em pair} $({\cal I}_A, {\cal P}_A)$ does not allow for universal control over the whole state-space ${\cal H}$
it is natural to find out whether there are subspaces ${\cal C}_i \subset {\cal H}$ such that:
\be
{\cal U}_A|_{{\cal C}_i}={\cal U}({\cal C}_i).
\label{encoded}
\ee
In this case we say that we have {\em encoded universality}, or that ${\cal U}_A$  is ${\cal C}_i$-universal.
The subspaces ${\cal C}_i$'s are referred to as codes, in the framework of quantum information theory these are the subspaces 
in which one can encode the information to be processed.

It is interesting to note that the  ${\cal C}_i$'s need not to be invariant subspaces for  ${\cal I}_A$, in fact the encoded 
universality can be achieved even by allowing the state of the system to temporarily leave the code subspace during 
the time evolution. A simple example of this situation is given by the subspaces of a code  ${\cal C}$: if ${\cal U}_A$  is 
${\cal C}$-universal than it is universal over all the subspaces   ${\cal C}' \subset {\cal C}$. In this case the evolution
starts and ends in a given subspace, althought  other intermediate states ({\em ancill\ae}) that belong to 
$({\cal C}^\prime)^\perp$ might be reached during the time evolution of the system.

It is  clear that, from the point of view of both  quantum control theory and quantum computation theory, it is fundamental
to find a general way to define which kind of control one has over a given quantum system.
In the following section we give a general procedure to address this problem.

\section{ General framework}
The idea is to find out under which conditions the physically available interactions allow to perform universal control over
the system $S$ or, if this is not possible, to give a procedure to identify the codes over which one can have encoded universality.
We start by giving a concise description of the mathematical ingredients involved and at the same time giving a 
first physical  example of their realization.

The example is given by a three-levels system which is
well known in atom-lasers physics . The Hilbert space of the system is given by 
$\mbox{span}\{\ket{1},\ket{2},\ket{3},\ket{e}\}$ where the first three states are the degenerate (atomic) ground states,
$\ket{e}$ is the first exited state; $\Delta E$ is the energy difference
between the excited and the ground state. The system is coupled with controllable external potentials which are lasers
characterized by the Rabi frequencies $\Omega_1,\Omega_2,\Omega_3 \in \CC$. Since the lasers
are detuned with respect to the energy gap $\Delta E$, the ground states are coupled to one another by a second order process,
but they are never coupled to the excited state. The Hamiltonian of the system can be written as
\be
H=\Omega_1 \ket{1} \bra{2}+\Omega_2 \ket{2} \bra{3}+\Omega_3 \ket{3}\bra{1}+h.c.
\label{ham3}
\ee
and the evolution of the system takes place in the degenerate ground state subspace, i.e. ${\cal H}=\mbox{span}\{\ket{1},\ket{2},\ket{3}\}$.

We now start by giving the basic ingredients of our control picture:
\begin{itemize}
\item{}
the system $S$, whith state-space ${\cal H}$, and the corresponding group of unitary operators ${\cal U}({\cal H})$;
\item{}
the set of physically available and controllable interactions ${\cal I}_A:= \{\sum_i \lambda_i H_i\};$
\item{}
the Lie algebra ${\cal L}_A$ generated from  ${\cal I}_A:= \{\sum_i \lambda_i H_i\};$  by commutation; 
\item{}
a discrete (continuos) group ${\cal K}$ and its (irreducible) representations;
\item{}
the group algebra $\CC{\cal K}$ generated by ${\cal K}$;
\item{}
the unitary part of the group algebra ${\cal U}\CC{\cal K}$.
\end{itemize}

In our first example we have already seen that ${\cal H}=\mbox{span}\{\ket{1},\ket{2},\ket{3}\}$, thus the unitaries over ${\cal H}$ belong to ${\cal U}(3)$. Since we suppose to be able to control the three lasers separately,
the set of avalable interactions is given by:
\[
{\cal I}_A:= \{\Omega_1 \ket{1} \bra{2}+h.c., \Omega_2 \ket{2} \bra{3}+h.c.,\Omega_3 \ket{3}\bra{1}+h.c.\}.
\]

Suppose now that ${\cal K}$  is a discrete group. We start by giving a description
of the group algebra $\CC{\cal K}$ \cite{isham},\cite{hamer}. This is defined as
\be
\CC{\cal K}=\{ x \ |\ x= \sum_{g\in{\cal K}} \lambda_g g \ ; \ \lambda_g \in \CC \},
\label{gralg}
\ee
that is, it is the vector space of all linear complex combinations of  elements of ${\cal K}$.
Its dimension is $\mbox{dim}_{\CC{\cal K}}=|{\cal K}|$, i.e. the number of elements of ${\cal K}$.
In order to complete the definition of the group algebra one has to define: i) a binary operation
(product) on it, ii) a unary operation (conjugation). The product law can naturally be defined 
by means of the group multiplication law; if $x,y \in \CC{\cal K}$, that is $x= \sum_{g\in{\cal K}} \lambda_g g$
and $y= \sum_{h\in{\cal K}} \lambda_h  h $, then:
\[
xy =  \sum_{g,h\in{\cal K}} \lambda_g\lambda_h  g h
\]
belongs to $\CC{\cal K}$  since $gh \in {\cal K}, \forall g,h $ . The conjugation is simply
defined $\forall x \in \CC{\cal K} $ as:
\[
x^*=\sum_{g\in{\cal K}} \bar \lambda_g g^{-1}
\]
where $\bar \lambda_g$ is the complex conjugate of $\lambda_g$, and $g^{-1}$ is the inverse of
$g$ with respect to the group law.

We will be interested in the representation of the elements of ${\cal K}$ as operators over a Hilbert 
space ${\cal H}$, i.e.  $\rho \ : \ {\cal K} \rightarrow {\cal L}({\cal H})$ a linear map onto the group of linear
operators on ${\cal H}$. It is interesting to notice that by means of the representation $\rho$ one can 
give a representation of $\CC{\cal K}$ over ${\cal H}$. In fact  $\forall x \in \CC{\cal K} $  the operator 
 $\rho(x)= \sum_{g\in{\cal K}} \lambda_g \rho(g)$ belongs to ${\cal L}({\cal H})$.
Thus, if ${\cal H}$ has finite dimension, from now on we can think of  the elements of ${\cal K}$ 
and  $\CC{\cal K}$  as complex $n \times n$ complex  matrices, where $n=\mbox{dim}{\cal H}$.

In our example we can choose the group ${\cal K}={\cal D}_3$, i.e.the group of rigid symmetries of a triangle.
We have that
\be
D_3 =\{ \openone, R, P, RP,PR,RPR\},
\ee
that is ${\cal D}_3$ has six elements and it can be written in terms of the 
two generators $R$ and $P$ ,which represent a rotation of $2\pi/3$ and a reflection, respectively. If we now represent the vertices of the triangle with the basis states $\ket{1}, \ket{2},\ket{3}$ we see 
that we can give a three dimensional representation $\rho$ of $D_3$ by writing $R$ and $P$ in terms of operators 
over  ${\cal H}=\mbox{span}\{\ket{1},\ket{2},\ket{3}\}$ in the following way: 
\be
\rho(R) = \ket{1}\bra{2}+\ket{2}\bra{3}+\ket{3}\bra{1};\ \  \rho(P) =\ket{1}\bra{2}+\ket{2}\bra{1}.
\label{repD3}
\ee 
By means of the same representation we can give a representation of  $\CC{\cal D}_3$, so that we can think
about the elements of $D_3$ and  $\CC{\cal D}_3$ as $3 \times 3$ complex matrices.

It is now interesting to have  a closer look to some of the elements that compose $\CC{\cal K}$. 
In order to do this we first notice a couple of nearly obvious facts: 
 i) if $x,y \in \CC{\cal K}$ then also $xy-yx$ belongs to $\CC{\cal K}$,i.e. the commutator of two elements of $\CC{\cal K}$
is again an element of the group algebra; ii) if we consider unitary representations of ${\cal K}$, $\rho \ : \ {\cal K} \rightarrow {\cal U}({\cal H})$, as it is in the case of \ref{repD3}, then the representative of the conjugate of an element of $\CC{\cal K}$
corresponds to the Hermitean conjugate of the representative of the element itself i.e., $\rho(x^*)=\rho(x)^\dagger$.

The first subset of $\CC{\cal K}$ that is fundamental for universal control is the unitary 
part of $\CC{\cal K}$ defined as: 
\[
{\cal U}(\CC{\cal K}) = \{x \in \CC{\cal K} \ | xx^* = x^* x = \openone \}; 
\] 
that is it is the set of elements of $\CC{\cal K}$ that are represented by $\rho$ as unitary operators over ${\cal H}$.
It is easy to verify, using the multiplication law of ${\cal K}$, that this set is actually a group.  
Since we suppose $\rho$ to be a unitary representation of ${\cal K}$  we have  $\rho({\cal K}) \subset \rho({\cal U}(\CC{\cal K})).$ 
The other important subset of $\CC{\cal K}$ is the set of all the anti-hermitian operators: 
\[
u(\CC{\cal K}) = \{x \in \CC{\cal K} \ | \ x = -x^* \}
\]
this is actually a Lie algebra, furthermore we have that ${\cal U}(\CC{\cal K}) = \exp(u(\CC{\cal K}))$, 
that is the unitary elements of the 
group algebra can be built out of the elements of $u(\CC{\cal K})$ via exponentiation.
In our example we have that 
\be
u(\CC{\cal D}_3) = i\alpha P - \beta R + \bar \beta R^*,
\label{hermD3}
\ee
with $\alpha \in \R, \beta \in \CC$

We now give some standard results of the theory of groups representation that will be useful in the following paragraphs. 
Suppose $\rho$ is a representation of the group ${\cal K}$ on ${\cal H}$. Then  $\rho$ is a reducible representation
if there is at least one subspace ${\cal H}' \subset {\cal H}$ that is invariant with respect to the action of the 
elments of ${\cal K}$, that is ${\cal H}'$ is such that $\forall \ x \in {\cal K}$ and $\forall \ \ket{\phi} \in {\cal H}'$ 
we have that $\rho(x)\ket{\phi} \in {\cal H}'$. In this case the restriction $\rho_{|{\cal H}'}$ is itself
a representation of  ${\cal K}$ on  ${\cal H}'$. If the only invariant subspaces of ${\cal H}$ with respect to $\rho$
are ${\cal H}$ itself and $\{0\}$ then the representation is called irreducible.

In general, if  $\rho$ is reducible it induces on
${\cal H}$ the following decomposition: 
\be
{\cal H}=\oplus_{J\in {\cal I}} {\cal H}_J,
\label{deco}
\ee
that is ${\cal H}$ can be written as the direct sum of the  orhotogonal invariant subspaces ${\cal H}_J$,  
on which the group acts irreducibly (${\cal I}$ is the set of 
indexes that label the various subspaces). Thus the restriction  $\rho_{|{\cal H}_J}$ of the representation to 
any of these subspaces is an irreducible representation of  ${\cal K}$ that has dimension $d_J = \mbox{dim}{\cal H}_J$. 
Notice that in Eq. (\ref{deco}) the sum is over possibly {\em equivalent} irreps. If the label $J$
is meant to represent just inequivalent irreps, then  one has to introduce factors $\CC^{n_J}$
taking into account the multiplicity $n_J$ with which the $J$-th irrep appears \cite{ZL-ctrl}.

In the case of the three-level system, it can be easily checked that, given the representation (\ref{repD3}) of $D_3$,
the Hilbert space ${\cal H}=\mbox{span}\{\ket{1},\ket{2},\ket{3}\}$ can be written as ${\cal H}={\cal H}_1 \oplus {\cal H}_2$,
where 
\be
{\cal H}_1 = \mbox{span}\{ ( \ket{1}+\ket{2}+\ket{3})/\sqrt 3 \},
\label{H1}
\ee
\be
{\cal H}_2 = \mbox{span}\{ \sum_{j=1}^3 \exp(i2\pi j/3 ) \ket{j},\sum_{j=1}^3 \exp(i4\pi j/3 ) \ket{j} \}.
\label{H2}
\ee
Thus, ${\cal H}_1$ and ${\cal H}_2$  are invariant subspaces, and we have that 
the restriction $\rho_{|{\cal H}_1}$ ($\rho_{|{\cal H}_2}$) of (\ref{repD3}) to ${\cal H}_1$ (${\cal H}_2$)
gives a one (two) dimensional representation of $D_3$.

We can now give the result of representation theory that is fundamental for the universal control.
Suppose  $\rho$ is an irreducible representation of  ${\cal K}$ over ${\cal H}$, then, with respect to 
the same representation, 
\be
{\cal U}(\CC{\cal K}) = {\cal U}({\cal H}).
\label{prop}
\ee
This means that 
if, with the available and controllable interactions ${\cal I}_A$, we are able to generate 
by commutation (see (\ref{trot1}) and (\ref{trot2}))  all the hermitian elements of $u(\CC{\cal K})$ (Hamiltonians), then, since 
${\cal U}(\CC{\cal K}) = exp(u(\CC{\cal K})))={\cal U}({\cal H})$,
we can have the universal control over the system $S$.

The result can be extended to the case in which $\rho$ is a reducible representation of  ${\cal K}$ over ${\cal H}$. In this case for any irreducible subspace ${\cal H}_J$ of the decomposition (\ref{deco}),
we have that 
\be
{\cal U}(\CC{\cal K})_{|{\cal H}_J} ={\cal U}({\cal H}_J).
\ee
This means that 
if, with the available and controllable interactions  ${\cal I}_A$ , we are 
able to generate by commutation (see (\ref{trot1}) and (\ref{trot2})) all the anti-hermitian elements  of 
$u(\CC{\cal K})$, then we can generate all the unitary operators over 
each ${\cal H}_J$ and so we can have encoded universality over each of the irreducible subspaces.

In the case of the three-level system we have that  ${\cal H}_2$ is an invariant subspace for the representation 
(\ref{repD3}) of $D_3$, and the restriction $\rho_{|{\cal H}_2}$ of (\ref{repD3}) to ${\cal H}_2$ is irreducible.
This means that, since with the available interactions we are able to generate the whole $u(\CC{\cal D}_3)$,
see (\ref{hermD3}), we have that  ${\cal U}(\CC{\cal D}_3)_{|{\cal H}_2} ={\cal U}({\cal H}_2)$, i.e. we 
have encoded universality over the code ${\cal C}_2={\cal H}_2$ (and the same is obviously true for the one 
dimensional invariant subspace ${\cal H}_1$).

%Now that we have the complete framework we can think of at least two different ways in which the result can be used. 
We are now in the position to summarize  our conceptual framework for universal quantum control.

The basic ingredient  is the set of physical interactions ${\cal I}_A$ 
available and controllable for a given system $S$.
In order to check if the controllable interactions allow to have universal
control over  ${\cal H}$, one has to:
\begin{itemize}
\item{}
find a group  ${\cal K}$ and a representation $\rho$ over ${\cal H}$;
\item {}
find the decomposition ${\cal H}=\oplus_J {\cal H}_J$ with respect to $\rho$; 
\item{}
check if the set of available interactions ${\cal I}_A$ allow, via 
commutation  (see (\ref{trot1}) and (\ref{trot2})) to generate the whole Lie algebra $u(\CC{\cal K})$. 
\end{itemize}
Then we have two cases: $i)$  $\rho$ is irreducible: then one has universal control over the whole state-space
${\cal H}$; $ii)$  $\rho$ is reducible: then one can achieve encoded universality over all the irreducible subspaces
${\cal H}_J$ (codes) of the decomposition (\ref{deco}). 

A first important example for the group ${\cal K}$ is
\be
{\cal K}={\cal U}_A=\exp( {\cal L}_A ).
\ee
In order
to find out if it is possible to have  encoded universality, one has to look at the irreducible representations 
of ${\cal U}_A$. In this case one thing should be clear: although ${\cal U}_A$ acts irreducibily on each of the 
subspaces ${\cal H}_J$ of the decomposition (\ref{deco}) this does {\em not} imply that 
${\cal U}(\CC{\cal U}_A)_{|{\cal H}_J} ={\cal U}({\cal H}_J)$. In fact, to have encoded universality in the irreducible
subspaces one has to prove that
\be
{\cal L}_A = u(\CC {\cal U}_A ),
\ee
i.e. the Lie algebra generated from the set ${\cal I}_A$  via commutation coincides with the Lie algebra of all
the anti-hermitian elements belonging to the group algebra ${\cal U}(\CC{\cal U}_A)$.

Before concluding this section it is worthwhile  to stress that the dynamical-algebraic machinery we developed allows one to
sistematically build classes of (to begin with) formal examples and then to look whether they admit some natural physically
realization.  The way  to proceed in this case  is:

\noindent i) choose a group  ${\cal K}$ and a representation $\rho$ over an Hilbert 
space ${\cal H}$;

\noindent ii)
find the decomposition ${\cal H}=\oplus_J {\cal H}_J$ with respect to $\rho$; 

\noindent iii)
find the minimal set of interactions ${\cal I}_A$ that allow via commuation (see (\ref{trot1}) and (\ref{trot2})),
to generate the whole Lie algebra $u(\CC{\cal K})$;

\noindent iv)
find a physical system whose state space is isomorphic to ${\cal H}$ and for which one has the
control of the minimal set of interactions ${\cal I}_A$.

\noindent Then, for the given physical system, one can generate all the unitary evolutions for each of the subspaces ${\cal H}_J$.

In the next sections we will analyze some physical models and show how the general theory applies to them.
In Ref. \cite{ZL-ctrl} the emphasis was mostly on spin $1/2$ systems i.e., qubits,
with reducible action of the available evolutions group ${\cal U}_A.$
Here we will be mainly concerned with different kind of bosonic systems relevant to   quantum information processing. 
For those systems the group  of available evolutions will have an irreducible action over
the physical state-space. Accordingly those examples cannot be regarded, strictly speaking, as instances of encoded universality. 
 
\section{ Bosonic double-well}
In this section we focus on a well known and well studied model: the Bose-Hubbard (BH) model for two boson wells
filled with a fixed number of bosonic particles $N$ \cite{milb_twowells}. In this model, the two wells are given by an external potential appropriately shaped. The particles occupy the lowest energy levels in each site and can be allowed, by lowering the barrier between the
wells, to tunnel from one site to the other. We suppose that the offset of the ground state of a site 
with respect to the other can be controlled by applying an additional external potential. Furthermore, the particles can interact
among each other when they are in the same site via a two-body scattering process (self-interaction).
The BH model has been recently used to investigate many interesting properties of different systems going from arrays of BEC where the
sites form an infinite (periodic) lattice \cite{bec_lattice}, to systems in which the number of sites is finite 
(dimer, trimer,graph structures) \cite{giorda_zanardi}). Moreover, this model has recently been used to describe new 
possible schemes for quantum computing \cite{radu_pz}. According to this model the Hamiltonian of the system 
composed of two  sites filled with $N$ particles can be written as:
\be
H = \gamma_1 n_1 + \gamma_2 n_2 +\tau (c_1^\dagger c_2 +c_2^\dagger c_1 )+ \varepsilon [n_1( n_1- 1)+n_2 (n_2- 1)], 
\label{twobosham}
\ee 
where: the indexes $1,2$ label the sites; $c_i^\dagger$ and $c_i$ ($[c_i,c_i^\dagger]= \openone$) are 
bosonic creation and annihilation operators that create/annihilate  particles in the local well $i$; $n_i=c_i^\dagger c_i$
is the corresponding occupation number operator. The parameters $\gamma_i$ can be used 
to model the offsets in the ground state energies of the different sites.. The parameter $\varepsilon$ controls the nonlinear two-body 
interaction between the particles in each site, while the parameter $\tau$ control the tunneling processes between the sites. 
In general the system lives in the Fock space ${\cal H}_F =h^{\otimes L}$, where $h = \{\ket{n}\}_{n=0,..,\infty}$ is the state space of a 
single quantum harmonic oscillator and $L = 2$ is the number of bosonic wells (sites). 
If the number of particles $N$ is fixed, as we shall assume, the corresponding Hilbert space is 
\be
{\cal H}_N = \mbox{span}\{\ket{N-n}_1 \ket{n}_2 \}_{n=0,..,N}, 
\label{HilN}
\ee 
where $\ket{N-n}_1 \ket{n}_2$ is the basis state corresponding to the situation in which the are $N-n$ bosons
in the first well and $n$ bosons in the second one. ${\cal H}_N$ is a subspace of ${\cal H}_F$ with 
dimension $d_{N,L}:=\pmatrix{&N+L-1\cr &L-1}$. 

In order to study the control properties of the system it is useful to resort to the Schwinger picture of spin operators 
\cite{saku}, by means of  which it is possible to give the following two-bosons realization of the spin operators $X,Y,Z$: 
\[
X = c_1^\dagger c_2 +c_2^\dagger c_1; \ Y=i(c_2^\dagger c_1 - c_1^\dagger c_2); \ Z = \frac{n_1- n_2}{2}. 
\] 
These operators are the generators of the Lie algebra $su(2)$ and satisfy the commutation relation $[A,B] = i\epsilon_{ABC}C$, 
where $A, B, C \in \{X,Y,Z\}$ and $\epsilon_{ABC}$ is the totally antysimmetric tensor.
In this representation the Hamiltonian (\ref{twobosham}) can be rewritten as: 
\be
H = \gamma_A Z +\tau X + 2\varepsilon Z^2 + \gamma_S N + \varepsilon N^2/2,
\label{HamSchw}
\ee 
where $\gamma_A=(\gamma_1-\gamma_2)$,\  $\gamma_S=(\gamma_1+\gamma_2-2\varepsilon)/2$.
\ $N = n_1 +n_2$ is the total number of particles operator and, as long as the 
number of particles in the wells is fixed, it is proportional to the identity operator. 

If we now suppose to be able to control the parameters $\gamma_A$, the offset between the ground states of the two wells,
$\tau$, the tunneling rate of the particles and $\varepsilon$, the value of the self-interaction among the particles
in the wells, we see from (\ref{HamSchw}) that our set of available interactions is ${\cal I}_A= \{X,Z, Z^2\}$. 

With this picture in mind, the group ${\cal K}$ that naturally arises is $SU(2) = \exp[su(2)]$.
The irreducible representations of this group are labeled by the index $J \in \{0, 1/2, 1, 3/2, 2..\}$ and the 
corresponding operators act on Hilbert spaces $H_J$ of dimension $d_J = 2J +1$. If we now suppose that the number of
particles in our sistem is fixed to $N$ we see that the corresponding Hilbert space ${\cal H}_N$ is isomorphic to $H_J$ 
with $J = (d_{ N,2}- 1)/2$.

Since $SU(2)$ is a Lie group the Lie algebra corresponding to the unitary part of the group algebra $\CC SU(2)$  
is  the universal enveloping algebra of $su(2)$. The latter is the vector space {span}ned by 
$\{A^a B^b C^c\}_{ a,b,c=1,..,\infty} \ ( A \neq B \neq C \in \{X, Y, Z\})$, i.e., is the vector space of 
the polynomials of any order in the $su(2)$ generators $X,Y, Z$ endowed with usual Lie bracket.
Thus, in order to prove that the set of available interactions allow us to have universal control over each of the ${\cal H}_N$'s,
we have to prove that by controlling the set of interactions ${\cal I}_A= \{X,Z, Z^2\}$ we can generate, by commutation
the whole universal enveloping algebra of $su(2)$. It turns out that this is true and it will be proved in  the next section. 
We can then say that we can generate the whole set of unitary operators ${\cal U} (H_N)$ 
(\ref{prop}).

Once that the possibility to achieve universal control over the ${\cal H}_N$'s
has been proven, one has to give for the desired value of $N$ an explicit representation of the operators that form a basis for the
$J = (d_{ N,2}- 1)/2=N/2$ representation of $su(2)$. 

We work out explicitly the simplest example that correspond to $N=2$. 
In this case we have that ${\cal H}_{N=2}$ has dimension $d_{2,2}=d_{J=1}=3$.
The whole group of unitary operators is then ${\cal U}(3)=\exp(u(3))$ and the canonical basis for $u(3)$ is given by the following $9$ elements:
\bae
X_{nm} &=& \ket{n}\bra{m}+ \ket{m}\bra{n},\ \ \ \ n<m \nonumber \\
Y_{nm} &=& i(\ket{n}\bra{m}- \ket{m}\bra{n}),\ \ \ \ n<m \nonumber \\
E_{nn} &=& \ket{n}\bra{n}. \nonumber \\
\eae
where $n,m \in \{0,1,2\}.$ In practice, in order to prove that is possible to generate all the unitary operators it sufficient to find a basis for the Lie algebra $su(3)$ which is the algebra of the $3 \cdot 3$ complex anti-hermitian matrices with trace  zero, whose dimension is $3^2-1$ \cite{stern},\cite{corn}.
Below we write a possible set of elements of the enveloping algebra of $su(2)$ that form this basis, 
each expressed in terms of the basis elements of $u(3)$ 
\bae
X &=& \sqrt 2 (X_{01}+X_{12}) \nonumber \\
Y&=& \sqrt 2 (Y_{01}+Y_{12})\nonumber \\
Z&=& 2(E_{22}-E_{00})\nonumber \\
X^2-Y^2 &=& 4X_{02}\nonumber \\
X^2-Z^2 &=& 2(X_{01}+ 2E_{11}-E_{00}-E_{22})\nonumber \\
XY+YX &=& 4Y_{02}\nonumber \\
ZX+XZ &=& 2\sqrt 2 (X_{12}-X_{01})\nonumber \\
ZY+YZ &=& 2\sqrt 2 (Y_{12}-Y_{01}),\nonumber \\
\eae
These relations show explicitly that by controlling
elements of the eneveloping algebra of the bosonic $su(2)$ up to degree two,
universality is achieved in the two-boson sector.  

\section{Enveloping algebra  of su(2)}
This section has a prevailing technical nature. The reader more interested in following  the  conceptual
stream of the paper may want to skip it at a first reading.
The goal here is to prove that it is possible to generate all the enveloping algebra of $su(2)$ by starting form the given controllable
interactions, i.e. ${\cal I}_A=\{X,Z,Z^2 \}$. The procedure is the following: $i)$ generate all the elements of order 
$\leq 2$; $ii)$ generate the elements of higher order by using a standard procedure that allows, given all the
elements of the enveloping algebra of $su(2)$ of order $\chi$ , to generate all the elements of order $\chi + 1$.
We first describe how to obtain the elements  of order $\chi\leq 2$. Obviously $Y=-i[Z,X]$. The term ZY is obtained 
from 
\be
[Z^2,X] = i(ZY + YZ)=2 i Z Y - X.
\label{zy}
\ee
 This first example allows us to describe an obvious an general rule that will be used extensively in the general procedure described below. In general the result of the commutation is not in the desired form,
for example the elements $X,Y,Z$ compare with the required power but they are not in the required order. The ordering 
process requires the use of the commutation relations of $su(2)$; these allow to write the starting element, $ZY$ in \ref{zy},
in terms of a new element of the same order  in which the position of two factors has changed, 
and of a new term of lower order $(X)$. By subtracting the lower order term from the result of the commutator we get the 
desired one $(YZ)$. In the same way we can write the terms $ZX \mbox{ and } XZ$.
The next step is to obtain the terms $X^2 \mbox{ and }Y^2$. This can be done by respectively adding and subtracting $iZ^2$ 
to the result of the commutators: 
\be
[Y,ZX]=-i(Z^2-X^2); \ \ \ [X,ZY]= i(Z^2-Y^2).
\ee
The remaining terms $XY \mbox{ and } YX$ can be obtained, for example, by  applying to  $[X^2,Z]$ the  same procedure 
of ordering and subtraction seen for (\ref{zy}). 

\vspace{0.5cm}

We now focus on the element of order $> 2$.
In order to describe the general procedure we use the operators $A,B,C \in \{X,Y,Z\}$ such that $A \neq B \neq C$.
A first obvious basic rule that can be applied to generate new elements of the enveloping algebra is to properly  commute elements already available 
at a certain step of the procedure. In general, the result of the commutation will be a linear combination of the desired new
element of order $\chi + 1$ with elements of the same or of a lower order. Thus, the new element will be obtained simply by 
applying an ordering process on terms of order $\chi + 1$ and by subtracting the resulting lower order terms.
Another important part of the procedure will entail the use of the Casimir relation that is satisfied by the elements
of $su(2)$:
\be
J(J+1) = A^2+B^2+C^2,
\label{casimir}
\ee
where $J$ is the label of the representation. This relation will be used to build terms of order $\chi + 1$ out of terms
of the same order that are already available.

A usful  relation that will be extensively used is the following;
let $O$ be a generic operator, then:
\be
[A^k,O]=\sum_{p,q\geq 0, p+q=k-1}A^p[A,O]A^q.
\ee

\vspace{0.1cm}
We now describe the first step of the procedure with which we can obtain elements of the type $A^{\chi+1}$. We first write:
\begin{widetext}
\bae
 [A^{\chi-1}B,CA] \ - \  [A^{\chi-1}C,BA] &=&   is_{BC}A^{\chi+1}\  - \ is_{CB}A^{\chi+1}\  - \  is_{CA}A^{\chi-1}B^2 \   + 
\  is_{BA}A^{\chi-1}C^2  \nonumber \\
&-& \  is_{AB}(\sum_{p,q\geq 0, p+q=\chi-2}A^pCA^qAC) \  + \  is_{AC}(\sum_{p,q\geq 0, p+q=\chi-2}A^pBA^qAB)
\label{chi1}
\eae
\end{widetext}
where, since $A,B \mbox{ and }C$ satisfy the $su(2)$ commutation relations, we have that $s_{BC}=\mbox{sign}([B,C])=-s_{CB},\ s_{BA}=\mbox{sign}([B,A])=
-s_{AB},\mbox{ and } s_{AC}=\mbox{sign}([A,C]=-s_{CA}$.
In each of the sums in (\ref{chi1}), the various terms are of the same order $(\chi + 1)$ and they differ from each other only 
for the order in which the $A's$ and the $B's (C's)$ appear. Since we want to apply the 
Casimir relation we have to first apply  the ordering process discussed above to the terms in the sums to obtain:
\be
\sum_{p,q\geq 0, p+q=\chi-2}A^pBA^qAB= nA^{\chi-1}B^2 + p(\chi)
\ee
and
\be
\sum_{p,q\geq 0, p+q=\chi-2}A^pCA^qAC= nA^{\chi-1}C^2 + p(\chi)
\ee
where $n$ is the number of terms of each sum and $p(\chi)$ represents a polynomial of order $\chi$.
In this way eq. (\ref{chi1}) can be rewritten as:

%\be
%\begin{array}{lr}
%[A^{\chi-1}B,CA]-[A^{\chi-1}C,BA]= & i2s_{BC}A^{k+1} \ +\\
%  + \  i(n + 1)s_{BA}A^{k-1} (B^2 + C^2 ) & +p(\chi) 
%\end{array}
%\label{chi4}
%\ee

\begin{widetext}
\be
[A^{\chi-1}B,CA]-[A^{\chi-1}C,BA]=  i2s_{BC}A^{\chi+1} \ 
  + \  i(n + 1)s_{BA}A^{\chi-1} (B^2 + C^2 )  +p(\chi) 
\label{chi4}
\ee
\end{widetext}
We now use the Casimir relation (\ref{casimir}) to substitute $(B^2 + C^2)$ in (\ref{chi4}) and we finally obtain:
\begin{widetext}
\be
[A^{\chi-1}B,CA]-[A^{\chi-1}C,BA]=  i(3+n)s_{BC}A^{\chi+1} \  + \  i(n + 1)s_{BA}A^{\chi-1} (j+1)j  +p(\chi) 
\label{chi5}
\ee
\end{widetext}
Since by hypotesis we already have all the terms of order $\leq \chi$ we can use them to extract the desired
term $A^{\chi + 1}.$

\vspace{0.5cm}

The next step of the procedure is based on the computation of
commutators of the following form: 
\be
[A^{\chi -\gamma +1},B^{\gamma+1}] = is_{AB} \sum_{p+q=\chi -\gamma;t+s=\gamma}A^pB^tCB^sA^q, 
\ee
where $p,q,t,s \geq 0$ and $\gamma \in \{1,..,\chi - 1 \}$. The terms in the sum have all the same sign 
$s_{AB}$ and have order $\chi+ 1$. Each term of the sum can be obtained by converting all the others 
into it by mean of an ordering process and then by subtracting the resulting linear combination of elements 
of order $\chi$. By changing $\gamma \in \{1,..,\chi - 1 \}$  we can obtain all the terms in which the operator C appears just once; a prototype of these terms is
\be
A^{\chi-\gamma}B^\gamma C.
\label{chi6}
\ee
 
If we now start from these terms and repeatedly use the Casimir relation (\ref{casimir}) we can get all
the terms like
\be
A^{\chi+1-\gamma-l}B^\gamma C^l
\label{chi7}
\ee
where $l$ is an odd integer and now $\gamma \in \{2,..,\chi -l- 1 \}$.
In fact, suppose the desired term is $A^{\chi-2-\gamma}B^\gamma C^3$, then 
one can apply the Casimir relation to the sum of two appropriate terms of the form (\ref{chi6}) and obtain:
\bae
A^{\chi-2-\gamma}B^\gamma C(A^2 +B^2)& =& j(j+1)A^{\chi-2-\gamma}B^\gamma C - \nonumber \\
&-&A^{\chi-2-\gamma}B^\gamma C^3, 
\eae
where $j(j+1)A^{\chi-2-\gamma}B^\gamma C$ is of order $\chi -1$ and can then 
be canceled, just as the terms of order $\chi$ that would appear whenever an ordering process is applied to obtain the desired term. 
%Thus one can get all the terms like $A^{\chi+1-\gamma-l}B^\gamma C^l$ with $l$ %odd, 
%applying the procedure just described to terms in which the operator $C$ %appears raised to the power $l= 2.$

\vspace{0.5cm}
 
The next step of the procedure is based on the computation of the commutator $[A^p, A^\alpha B^\beta C]$
and on the use of the Casimir relation. By properly choosing the exponents $p,\alpha,\beta$
the result of the commutator is the sum of terms of order $\chi +1$ and can be written in the
following compact way:
\bae
[A^p,A^\alpha B^\beta C] &=&  ims_{AB} A^{\alpha+p-1}B^{\beta -1}C^2 \ +\nonumber  \\
 &+& \  ins_{AC} A^{\alpha +p - 1}B^{\beta+1} \  + \ p(\chi)
\label{chi9}
\eae
The term $ins_{AC} A^{\alpha +p - 1}B^{\beta+1}$ is the result of the ordering process 
on the $n$ terms in which $A$ appears $\alpha +p - 1$ times and $B$ 
appears $\beta+1$ times. Analogously $ims_{AB} A^{\alpha+p-1}B^{\beta -1}C^2$ is the 
result of the ordering process on the $m$ terms in which the operators $A, B \mbox{ and } C$ appear $\alpha+p-1$,
 \ $\beta -1 $ and $2$ times respectively. The term $p(\chi)$ takes into  account  the terms of order 
$\chi$ produced by these ordering processes.

The procedure is based on the following steps.
First we use the Casimir relation on $A^{\alpha +p - 1}B^{\beta+1}$ to get: 
\bae
A^{\alpha +p - 1}B^{\beta+1} &=& A^{\alpha +p - 1}B^{\beta-1}j(j+1)- \nonumber \\
&-& A^{\alpha +p - 1}B^{\beta-1}A^2 - A^{\alpha +p - 1}B^{\beta-1}C^2. \ \ 
\label{chi10} 
\eae 
Then if  we subsitute (\ref{chi10}) in (\ref{chi9}) we can  write: 
\bae
[A^p,A^\alpha B^\beta C] &=&  i(m+n)s_{AB} A^{\alpha+p-1}B^{\beta -1}C^2 \ +\nonumber  \\
 &+& \  ins_{AC} A^{\alpha +p - 1}B^{\beta-1}A^2 \  + \ p(\chi)
\label{chi11}
\eae
where we have taken into  account that $s_{AC} =-s_{AB}$.
We first note that for $\alpha +p - 1=\chi -1$ and $\beta=1$, since we already have $A^{\chi+1}$, 
it is possible to generate all the terms of the type $A^{\chi-1}C^2$ and, by simply exchanging the
role of $C$ and $B$, \  $A^{\chi-1}B^2$. 
If we now set $\alpha +p - 1=\chi -2$ and $\beta=2$, we can obtain the terms of the form 
$A^{\chi}B$ from (\ref{chi11}) and $A^{\chi- 2}B^3$ from (\ref{chi9}). 
In fact, in both cases all we have to do is to cancel out the terms like $A^{\chi-2}BC^2$ that has
been already obtained, see \ref{chi6}.

For $\beta \geq 3$ the procedure use at each step the results obtained 
in the previous ones, that correspond to lower values of $\beta$: 
first use (\ref{chi11}) to obtain the terms like 
\be
A^{\chi - \beta}B^{\beta -1}C^2,
\label{chi11zero}
\ee
then use 
(\ref{chi9}) to get the terms of the form 
\be
A^{\chi - \beta}B^{\beta +1}.
\label{chi11bis}
\ee

\vspace{0.5cm}

The last part of the procedure is basically the same used to obtained (\ref{chi7}); we start from the terms of 
the type $A^{\chi - \beta}B^{\beta -1}C^2$ and, by repetedly using the Casimir relation we can get all the terms of the type 
\be
A^{\chi+1-l-\gamma}B^\gamma  C^l
\label{chi11ter}
\ee
where now $l$ is an even integer. For example 
to obtain $A^{\chi-2-\gamma}B^\gamma  C^4$ we can write: 
\bae
 A^{\chi-3-\gamma}B^{\gamma}C^2 (A^2+B^2) &=&  A^{\chi-3-\gamma}B^{\gamma}C^2 j(j+1)- \nonumber  \\
 &-& \ A^{\chi-2-\gamma}B^\gamma  C^4
\label{chi12}
\eae
where the first term on the right hand side is of order $\chi-1$ 
and then can be canceled. 

\vspace{0.5cm}

If we now collect the results (\ref{chi5}),(\ref{chi6}),(\ref{chi7}),(\ref{chi11zero}),(\ref{chi11bis}) and (\ref{chi11ter}) we see that we have succeded in giving an operative procedure to build all the element of the enveloping algebra of $su(2)$,
starting from the set of available controllable interactions, i.e.,  ${\cal I}_A=\{X,Z,Z^2 \}$.

\section{ MASSLESS BOSON MODES} 
Another interesting physical example is given by a set of $L$  bosonic modes each representing 
the state space of massless bosonic particles. The quantum state-space in this case given by ${\cal H}_F =\otimes_{i=1}^L h_i$,
where $h = \{\ket{n}\}_{n=0,..,\infty}$. Independent modes of the electromagnetic field
e.g, photons with different polarization,
provide a prototypical and ubiquitous physical example of this kind of systems.

Each of single-mode Hilbert spaces $h_i$ is an irreducible representation of the Weyl-Heisenberg algebra
\begin{equation}
h(1)= \{c_i, c_i^\dagger, \openone\},\quad  [c_i,c_i^\dagger]=\openone,(i=1,\ldots,L).
\end{equation}
According to the general theory of section III, in order to achieve universal control over ${\cal H}_F$ one has to be able to 
generate, as in theformer $SU(2)$ case,
the wole Hermitean part of the enveloping algebra of $\otimes_{i=1}^L h_i(1)$ i.e., all the possible polynomials
in the $c_i$'s and $c_i^\dagger$'s. Notice that this enveloping algebra is now given in an infinite-dinsional   
representation, whereas in the former $SU(2)$ bosonic example the number superselection
allowed us to focus on finite-dimensional irreps (the Fock sectors).

It is a well known result in quantum control theory 
\cite{lloydbraun} that if is  able to control --for example for the  mode $i$-- 
the following set of interactions 
\be
{\cal I}_{A_i}= \{ c_i,c_i^\dagger,(c_i)^2,(c_i^\dagger)^2,n_i,(n_i)^2 \}
\label{masslessint}
\ee
then one has universal control on the mode-system, i.e., one is able to generate, via commutation,
linear combinations of Hermitian polynomials  in  $c_i \mbox{ and } c_i^\dagger$ of any degree.
The presence of the  non-linear tern  $(n_i)^2$ is essential, because it entails to increase via commuation
the order of the polynomials already available. 

Furthermore, if in addition to ${\cal I}_{A_i}$ one has the control over the set ${\cal I}_{A_j}$ and of two-modes interactions:
\[
{\cal I}_{A_ij}=\{ c_i^\dagger c_j,c_j^\dagger c_i\}
\]
then one has the   universal control over the whole two-modes system,i.e. one can generate, via
commuation, linear combinations of Hermitian polynomials  in  
$c_i,c_i^\dagger,c_j \mbox{ and } c_j^\dagger$ of any degree \cite{lloydbraun}.

What we first want to show here is that it is sufficient
to control the non-linear terms of only one of the bosonic modes, say  $(n_1)^2$, in order to have
universal control over each of the modes $i \in \{2,..,L\}$.
If we suppose to be able to control the following set of interactions:
\be
{\cal I}_{A_1}= \{ c_1,c_1^\dagger,(c_1)^2,(c_1^\dagger)^2,n_1,(n_1)^2 \},
\label{masslessint1}
\ee

\be
{\cal I}_{A_i}= \{ c_i,c_i^\dagger,(c_i)^2,(c_i^\dagger)^2,n_i \},\ \ \ i \in \{2,..,L\},
\label{masslessint2}
\ee
and
\be
{\cal I}_{A_ij}=\{ c_i^\dagger c_j,c_j^\dagger c_i\},\ \ \ i \neq j,\;  i,j \in \{1,..L\},
\label{masslessint3}
\ee
then we see that in order to reach our goal we have to prove that we are able to generate each of the terms $n_i^2,\ i \in \{2,..,L\}$.
This can be done by using the following commuators:
\be
[[n_1^2,T_{1i}],T_{1i}]= -8n_1n_i + 2c_1^2(c_i^\dagger)^2 + 2c_i^2(c_1^\dagger)^2+ P;
\label{nonlin1}
\ee
\be
[n_1,[n_1,[[n_1^2,T_{1i}],T_{1i}]]]= 8c_1^2(c_i^\dagger)^2 + 8c_i^2(c_1^\dagger)^2;
\label{nonlin2}
\ee
\be
[[n_1n_i, T_{1i}], T_{1i}]= n_2^2+ Q, 
\label{nonlin3}
\ee

where $T_{1i}= c_1^\dagger c_i + c_1 c_i^\dagger$ and $P,Q$ are polynomials in terms already available. 
First the procedure requires the use of (\ref{nonlin2}) in order to extract from (\ref{nonlin1}) the term $n_1n_i$. 
Then the latter is used in (\ref{nonlin3}) to generate the desired term $n_i^2$. We have thus shown that it is possible 
to "propagate" the non linearity $n_1^2$ to all the other modes. 

The control over the sets of interactions (\ref{masslessint}), (\ref{masslessint2}), (\ref{masslessint3}) and of the terms $n_i^2$ allow \cite{lloydbraun} universal control over any of the two-modes subspaces $h_i \otimes h_j$. In fact with the previous
 hypotesis, given a two-modes operator $O_i \otimes A_j$ one can always generate all the possible operators $O_i \otimes P_j$, where $P_j$ is a generic linear combination of Hermitian polynomials  in  $c_j \mbox{ and } c_j^\dagger$ of any degree; this can be done by judiciusly commuting $O_i \otimes A_j$ with operators $\openone \otimes B_j$ acting on the $j$'th mode only.
\noindent Analogous arguments can be used to generate multi-modes operators, and this means that the sets of interactions  (\ref{masslessint}), (\ref{masslessint2}), (\ref{masslessint3}) allow universal control over the whole $L$-modes state space ${\cal H}_F$.

\section{Auxiliary boson}

In this section we will discuss a technique for implementing effective Hamiltonians in a group-algebra $\CC{\cal K}$
which involves the use of an auxiliary bosonic mode. This an application of the geometric-phase operators discussed
in Refs \cite{Milburn1},\cite{Anders},\cite{Wangate},\cite{XZ}.

We consider a quantum state-space $\cal H$ which is an irrep space of the group ${\cal K}.$
The representatives of the group elements have the form $g=\exp(i H_g)\in\mbox{End}({\cal H})\,(H_g=H_g^\dagger).$ 
Our aim is to generate an effective dynamics governed by an Hamiltonian in $u\CC{\cal K},$ of the form $g+g^\dagger$ or $i(g-g^\dagger).$ 
These two operators, in terms of the Hamiltonians $H_g,$ are clearly  proportional to $\cos(H_g)$ and
$\sin(H_g)$ respectively.   Now the point is that it is known how to enact evolutions associated to this kind of
highly non-linear operators by resorting to a tunable coupling with an ancillary bosonic mode. This coupling has to 
be of the form  $K_g:=\theta H_g\otimes n$ where $n=a^\dagger a $ is the boson number operator ($[a, a^\dagger]=1$) 
Moreover one has to be able to control the bosonic mode itself in order to enact the  dynamics described by the 
{\em displacement}  operators
\begin{equation}
 D(\alpha):= \exp(\alpha a- \bar{\alpha} a^\dagger),\quad(\alpha\in\CC).
\label{disp}
\end{equation}
When one has at disposal such ingredients the following relation shows how to generate the desired effective
Hamiltonian over $\cal H$ while leaving the bosonic mode unaffected   
\begin{equation}
D(-\beta) D(-\alpha e^{i\theta H_g})D(\beta) D(\alpha e^{i\theta  H_g})=e^{i|\alpha||\beta| \sin(\theta H_g+\phi)}
\label{simul}
\end{equation}
Where $ D(\alpha e^{i\theta  H_g})= e^{i\theta n H_g}\,D(\alpha)\, e^{-i \theta H_g}$ and $\phi:=\arg(\alpha\bar{\beta}).$

For a finite order group $\cal K$ the most general element in u$\CC{\cal K}$ has the form
$\sum_g(\lambda_g \cos(H_g) +\mu_g \sin(H_g)),\,(\lambda_g,\mu_g\in\R).$ It follows that if one has available all the couplings $K_g$
the simulating sequence (\ref{simul})  allows to generate all possibile Hamiltonians in the group algebra of $\cal K$.
Even when just some, possibly just one, $K_g$ coupling is actually available on may still be able to generate the
whole u$\CC{\cal K}$, by generating commutators with other simple Hamiltonians at disposal.

In order to exemplify this latter point  let us consider the continous group $SU(2)$ and its irreps corresponding to 
total angular momentum $J.$
Since the irrep has dimension $2J+1$ one has e.g., that $\sin(\theta S^z)=\sum_{k=0}^{2J} F_J(\theta) (S^z)^k$ where
the $F_J$ are real-valued functions. The implementation of the trascendental  function $\sin$ in finite-dimensional spaces
 just amounts to generating a high order polynomial. From the previous analysis we know that this, along with the capability
to switch on and off simpler Hamiltonians e.g., $S^x,S^y,$ may be sufficient for generating all the Hamiltonians in $\CC{\cal K}.$
\section{conclusions}
In this paper we thoroughly analyzed the dynamical-algebraic approach to  universal quantum control  introduced 
in \cite{ZL-ctrl}. The basic idea is to decompose the quantum state-space $\cal H$ encoding information into irreducible sectors
and subsystems associated to the group of available evolutions. When this group coincides with the unitary part
of the group-algebra $\CC{\cal K}$ of some (non-abelian) group $\cal K$ then universal control is achievable over  the
${\cal K}$-irreducible components of $\cal H$. Physically this can be done by turning on and off a generic
pair  of Hamiltonians in $\CC{\cal K}.$  We then applied this general strategy to different kind of bosonic systems.
We first considered massive  bosons in a double-well and showed how to get universal control over all the finite-dimensional 
the Fock sectors by the control of single-well frequency, tunneling and a non-linear term. 
We then discussed the multi-mode massless  case showing again that linear 
(in the optics sense) Hamiltoniain along with a single non linearitity allows one
to generate the whole (infinite-dimensional) multi-mode Heisenberg-Weyl enveloping-algebra    
and so to achieve universality. 
Though some similar  results have been already derived in other situations it is remarkable the fact that they provide
a conceptually straightforward and enlighting exemplification of the general control-theoretic framework
introduced in Ref. \cite{ZL-ctrl}.  Notice that in those bosonic examples we gave a constructive way to generate 
effective interactions,  out of a simpler ones, by means of higher order commutators. 
This in turns provides a explicit simulation sequences for complex interactions e.g., high-order  non-linearities   
in terms of readily available ones. Finally we discussed how to use an auxiliary bosonic mode coupled with
finite-dimensional systems to generate the high-order non-linearities necessary to realize universal control.
We believe that the results  presented  here shed further light  over the  strengh and reach
of the universal quantum control approach introduced in \cite{ZL-ctrl}.

\acknowledgements{We thank R. Ionicioiu for a careful reading of the manuscript.
P.G. thanks the Mechanical Engineering Department of the MIT
for warm hospitality. 
P.Z. gratefully acknowledges financial support by
Cambridge-MIT Institute Limited and by the European Union project  TOPQIP
(Contract IST-2001-39215)}


\begin{references}

\bibitem{QC} For reviews, see D.P. DiVincenzo and C. Bennet  {\sl Nature} {\bf 404}, 247 (2000);
 A. Steane, Rep. Prog. Phys. {\bf 61}, 117 (1998)

\bibitem{SETH} S. Lloyd, Phys. Rev. Lett. {\bf 75}, 346 (1995)
\bibitem{UG} D. Deutsch, A. Barenco and A. Ekert, Proc. R. Soc. London { A}, {\bf 449}, 669 (1995);
D.P. Di Vincenzo, Phys. Rev. A, {\bf 50}, 1015 (1995)

\bibitem{mike}
J.~L.~Brylinski and R.~Brylinski, quant-ph/0108062; M.~J.~Bremner et al,  Phys. Rev. Lett. {\bf 89},
247902 (2003)


\bibitem{danFT}
D. Bacon, J. Kempe, D. A. Lidar, and K. B. Whaley
 Phys. Rev. Lett. {\bf 85}, 1758 (2000);
For a nice review see also D. Bacon, PhD Thesis, quant-ph/0305025
\bibitem{danNS}
 Kempe, D. Bacon, D.A. Lidar and K.B. Whaley, Phys. Rev. A {\bf 63}, 042307 (2001)

\bibitem{enc} D. P. Di Vincenzo et al, Nature {\bf 408}, 339 (2000)

\bibitem{Dan}  D. Lidar, L.-A  Wu, Phys. Rev. Lett. {\bf 88}, 017905 (2002);
L.-A. Wu and D.A. Lidar, {\em ibid}  88, 207902 (2002);
M. S. Byrd, D. Lidar, {\em ibid}  {\bf 89}, 047901 (2002)

\bibitem{lorenza} L. Viola,
Phys. Rev. A {\bf 66}, 012307 (2002)

\bibitem{ZL-ctrl} P. Zanardi, S. Lloyd quant-ph/0305013 
\bibitem{corn} 
J.F. Cornwell, {\em Group Theory in Physics}
(Academic, New York, 1984), Vol. I--III



\bibitem{EAC} L.M. Duan and G.C. Guo, Phys. Rev. Lett, {\bf 79}, 1953 (1997);
P. Zanardi and M. Rasetti, {\em ibid}  {\bf 79}, 3306 (1997);
D.A. Lidar, I.L. Chuang and K.B. Whaley, {\em ibid} {\bf 81}, 2594 (1998)


\bibitem{KLV} E. Knill, R. Laflamme and L. Viola, Phys. Rev. Lett. {\bf 84}, 2525 (2000);
L. Viola, E. Knill, S. Lloyd, {\em ibid}  {\bf 85}, 3520 (2000)

\bibitem{stab} P. Zanardi, Phys. Rev. A {\bf 63}, 12301 (2001)

\bibitem{lloydbraun} S. Lloyd and S. L. Braunstein, Phys. Rev. Lett. 82, 1784-1787 (1999)
\bibitem{stern} Sternberg, Group Theory and Physics, Cambridge University Press, 1994
\bibitem{milb_twowells} G.~J.~Milburn {\it et al.}, \pra {\bf 55}, 4318 (1997);
\bibitem{bec_lattice} D.~Jaksch {\it et al.}, \prl {\bf 81}, 3108 (1998).
\bibitem{giorda_zanardi} P.Giorda, P.Zanardi, quant-ph/0304151.

\bibitem{radu_pz} R.~Ionicioiu and P.~Zanardi, \pra 66, 050301 (2002)
\bibitem{saku} J.J. Sakurai, Modern quantum mechanics, Addison Wesley, 1985
\bibitem{isham} C. J. Isham, Lectures on groups and vector spaces for physicists, World Scientific, 1989.
\bibitem{hamer} M.Hamremesh, Group theory and its physical application to physical problems, Dover Pubblications, 1989

\bibitem{Milburn1}  G. J. Milburn, quant-ph/9908037.

\bibitem{Anders}  A. S\o rensen and K. M\o lmer, Phys. Rev. A {\bf 62},
022311 (2000).

\bibitem{Wangate}  X. Wang, A. S\o rensen and K. M\o lmer, Phys. Rev. Lett.
{\bf 86}, 3907 (2001).


\bibitem{XZ} X-G Wang, P. Zanardi, Phys. Rev A  {\bf 65}, 032327 (2002)

\end{references}
\end{document}